# Altitude and Latitude Distribution of Atmospheric Aerosol and Water Vapor from the Narrow-Band Lunar Eclipse Photometry

Oleg S. Ugolnikov [a,b,*] and Igor A. Maslov [a,c]

[a] *Space Research Institute, Profsoyuznaya st., 84/32, 117997, Moscow, Russia*
[b] *Astro-Space Center, Lebedev's Physical Institute, Profsoyuznaya st., 84/32, 117997, Moscow, Russia*
[c] *Sternberg Astronomical Institute, Universitetsky prosp., 13, 119992, Moscow, Russia*

**Abstract.** The work contains the description of two narrow IR-bands observational data of total lunar eclipse of March, 3, 2007, one- and two-dimension procedures of radiative transfer equation solution. The results of the procedure are the extinction values for atmospheric aerosol and water vapor at different altitudes in the troposphere along the Earth's terminator crossing North America, Arctic, Siberia and South-Eastern Asia. The altitude range and possible latitude and altitude resolution of atmosphere remote sensing by the lunar eclipses observation are fixed. The results of water vapor retrieval are compared with data of space experiment, the scale of vertical water vapor distribution is found.

## 1. Introduction

The lunar eclipses photometry is the only ground-based technique for atmosphere remote sensing with altitude resolution by measurements of tangent path absorption. During the eclipse the Moon crosses the Earth's umbra. Each point of this space is illuminated by solar emission refracted by the definite range of angles in the atmosphere of the Earth. The value of refraction angle is determined by the minimal altitude of the light ray path above the Earth's surface (or ray path perigee). The position angle of the point on the Moon determines location of the ray perigee above the Earth. Thus, different points of the umbra crossed by the Moon during the eclipse correspond to different perigee locations and different altitudes in the atmosphere. In fact, the scheme of the eclipse is similar to the one of space atmosphere absorption measurements, where the Sun is the light source and the Moon plays the role of spacecraft. The orbit radius of the Moon is sufficiently larger than the one of atmospheric measurement satellites. It causes both advantages and disadvantages.

    The advantage of the lunar eclipse method is slow dependency of effective ray perigee altitude on the position of the lunar surface point inside the umbra. It gives the possibility to expect high altitude resolution in the troposphere. Astronomical measurements technique brings high sensitivity which is important in the case of strong absorption along the troposphere tangent path. One more advantage of the method is large angular size of the Moon allowing to hold the simultaneous measurements in different parts of umbra. Each total eclipse the Moon covers the large part of northern (or southern) half of the umbra, giving the possibility to build the altitude distribution of atmosphere absorption along the northern (or southern) part of the Earth's limb – the line where the Sun and the Moon are at the horizon during the totality.

    The main problem of the method is large angular size of the Sun covering the definite range of refraction angles (or perigee altitudes) and position angles (or locations at the Earth's limb) being observed from the Moon. To convert the observational data to the resulting extinction values on the net of altitudes and locations, we have to solve the incorrect mathematical problem. It can be done using the regularization algorithms, restricting the horizontal and vertical resolution, and using a priori information about the possible solution values.

---

[*] Corresponding author. Fax: +7-495-333-5178. E-Mail: ougol@rambler.ru.

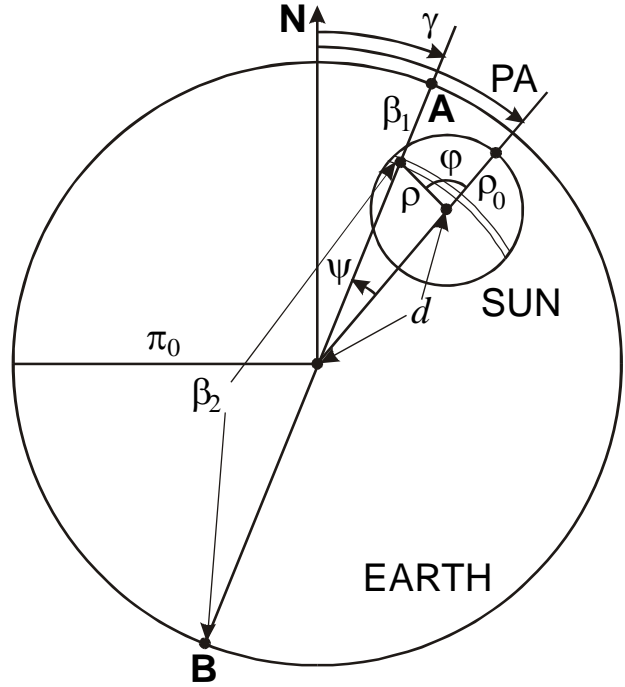

*Figure 1. Eclipse geometry being observed from the Moon.*

Wide theoretical and observational analysis of lunar eclipse phenomena was performed by Link in [1]. The theoretical part of this book contains the review of the methods of lunar surface brightness calculation during the eclipse. Observational part describes the correlation of eclipse magnitude (or brightness) with some characteristics that can influence on the magnitude, including the solar activity. As it was shown by A. Danjon in the early 1900s, the darkest eclipses (during those the Moon almost disappears from the sky) preferably happen right after the solar activity minimum. However, the reason of such correlation is not found, and the author did not reject it to be occasional.

The method of reverse problem solution was suggested in [2] for the observational data of two total lunar eclipses in 2004. The observations were carried out in two-peaked spectral band in red and near-IR ranges mostly avoiding the absorption bands of atmosphere gases. Comparing the observational data with gaseous atmosphere model calculations, authors found the additional extinction, which can be related with atmospheric aerosol. The reverse problem solution was one-dimensional, the solar disk reintegration was made only by the refraction angle $\beta$ (see Figure 1) basing on the dependency of brightness on the distance from the umbra center at definite umbra position angle *PA* (see Figure 2). The procedure works well if the latitude dependence of the aerosol extinction is slow. It also works in the outer umbra part with small refraction angle, where solar disk beyond the Earth covers small range of position angles $\psi$ (see Figure 1). These small refraction angles correspond to the altitudes about 10 km, where the aerosol distribution is compared with clouds cluster using the meteorological maps.

In this paper the two-dimensional method of reverse problem solution will be developed for the observational data of the total lunar eclipse of March, 3, 2007. The reintegration is done simultaneously by refraction angle $\beta$ and position angle $\psi$. The possible resolution and comparison with one-dimensional reintegration data are determining the possibilities of atmosphere remote sensing by the lunar eclipses observations. Measurements in two narrow spectral bands in the near-IR range outside and inside the $H_2O$ absorption lines allow to investigate not only aerosol, but also the water vapor distribution.

## 2. Observations.

Lunar surface photometry during the total eclipse of March, 3, 2007 was held at Southern Laboratory of Sternberg Astronomical Institute, Crimea, Ukraine (44.7°N, 34.0°E). The observations were carried out by two CCD-cameras: SBIG ST-6 with "Rubinar-500" lens (focal distance 500 mm, 1:8) and Sony DSI-Pro with "Jupiter-36B" lens (focal distance 250 mm, 1:3.5). Entire lunar disk was placed in both frames. Both devices worked with near-IR narrow-band interference filters with centering wavelengths equal to 867 and 938 nm and the width by half-maximum level equal to 10 and 55 nm, respectively. Glass color filters blocked the secondary maxima of interference filters. The exposure time varied in 100 times from non-eclipsed Moon to totality. First spectral interval is free from atmospheric gases absorption lines. The extinction in the atmosphere is formed by molecular and aerosol scattering. Last one has slow wavelength dependency and we can assume that aerosol extinction coefficient is the same for the wavelength 938 nm, where strong water vapor absorption is added. The effective cross-section of $H_2O$, integrated by the second observational band using the spectral measurements [3, 4] is equal to $5.6 \cdot 10^{-23}$ $cm^2$ $molecule^{-1}$.

Observations were carried out in variable weather conditions, the sky was partially cloudy. The atmosphere transparency at the observation point changed rapidly. To control it, the measurements of the star 56 VY Leonis in the same frame with the Moon were held. This star was situated close to the Moon during the whole eclipse, having experienced 3-min grazing occultation during totality. It is bright in near-IR enough to hold the CCD-photometry even near the eclipsed lunar disk and can be considered to be uniform light source in our bands on the time scale of several hours.

The methods of lunar surface photometry and background subtraction were analogous to [2]. The result of the procedure is the two-dimensional distribution of umbra darkening factor $U(d, PA)$. This value is equal to the ratio of brightness of the element of eclipsed lunar surface at distance $d$ from the center of umbra at the position angle $PA$ (see designations in the Figure 2) and the brightness of the same element outside the umbra and penumbra. This distribution inside the umbra is shown as the isophotes map in the Figure 2 for both wavelengths. The most noticeable feature visible in the both maps is the dark spot shifted far from the umbra center in equatorial direction. This minimum is also seen in the dependency $U(d)$ for position angle 90° (east equatorial direction) shown in the Figure 3 with the one for position angle 0° (northern polar direction) and theoretical curve for dry gaseous atmosphere model analogous to [2]. This spot can appear even in the case of monotonous dependence of the extinction coefficient on the altitude. Less brightness of umbra in the equatorial regions was also noticed in [2] for the other lunar eclipses and spectral passbands.

As it can be seen in the Figure 3, the brightness of eclipsed lunar disk is less than the one calculated for the dry gaseous atmosphere model. This difference is small in the outer regions and increases to the center. This picture is similar to the one observed for two eclipses in 2004 [2], and the eclipse of March, 3, 2007 was not darker than the previous ones as it was expected from the solar activity correlation [1]. We can also see that the umbra darkening factor in polar regions ($PA$ near zero) is principally the same for both wavelengths. It shows the vanishing level of water vapor in the northern winter troposphere. Equatorial part of umbra, including the dark spot, shows sufficient difference of umbra darkening factor for two wavelengths. This difference is related with high amount of water vapor in the equatorial troposphere. It will be proved numerically in the next chapter of this work.

**3. Two-dimensional reintegration procedure.**

Each point inside the umbra is illuminated by the emission of the Sun refracted in the atmosphere of the Earth. The Sun has large angular size (just about 3.5 times less than the one of Earth if we observe from the Moon). So the emission of different parts of the solar disk is refracted by different angles (at different altitudes) above the different points of the Earth's limb. To calculate the extinction at definite position angle and altitude, the reintegration procedure is needed. In paper [2] this procedure was one-dimensional, solar disk was divided into arcs with the same refraction angle and perigee altitudes, the data was related with the point on the limb corresponding to the middle of the arc (see Figure 1). Here the arcs are also divided into the zones corresponding to different limb locations. However, the one-dimensional reintegration procedure [2] will be also run for comparison of results.

The umbra darkening factor of the point at the angular distance $d$ from the umbra center and position angle $PA$ is equal to [2]:

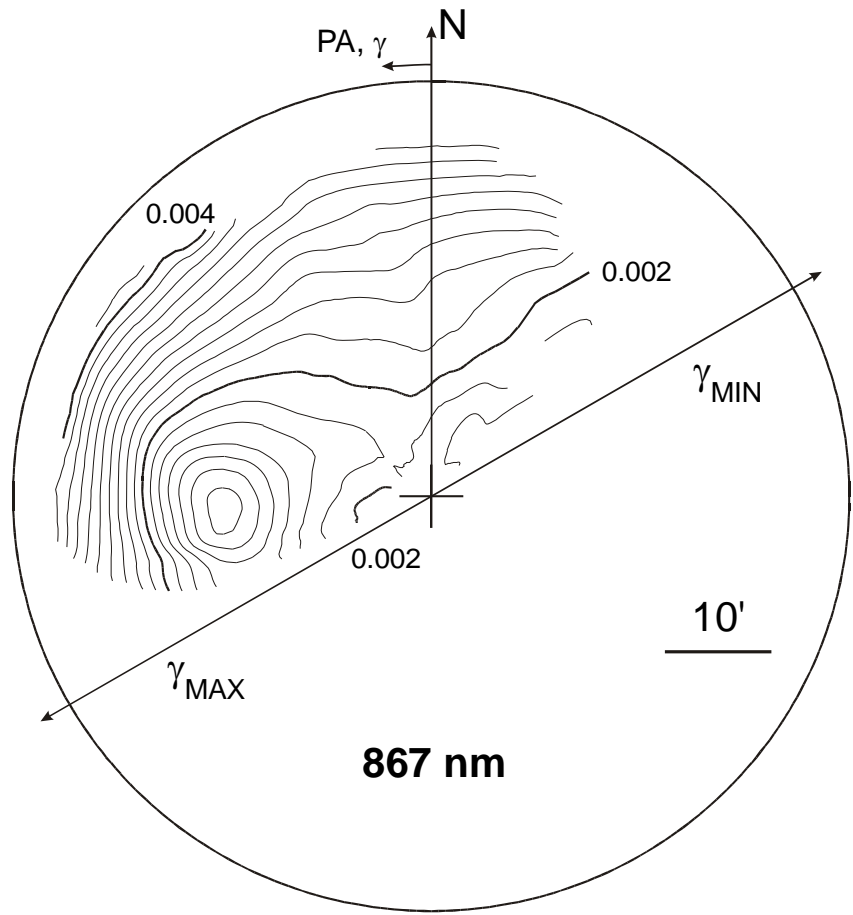

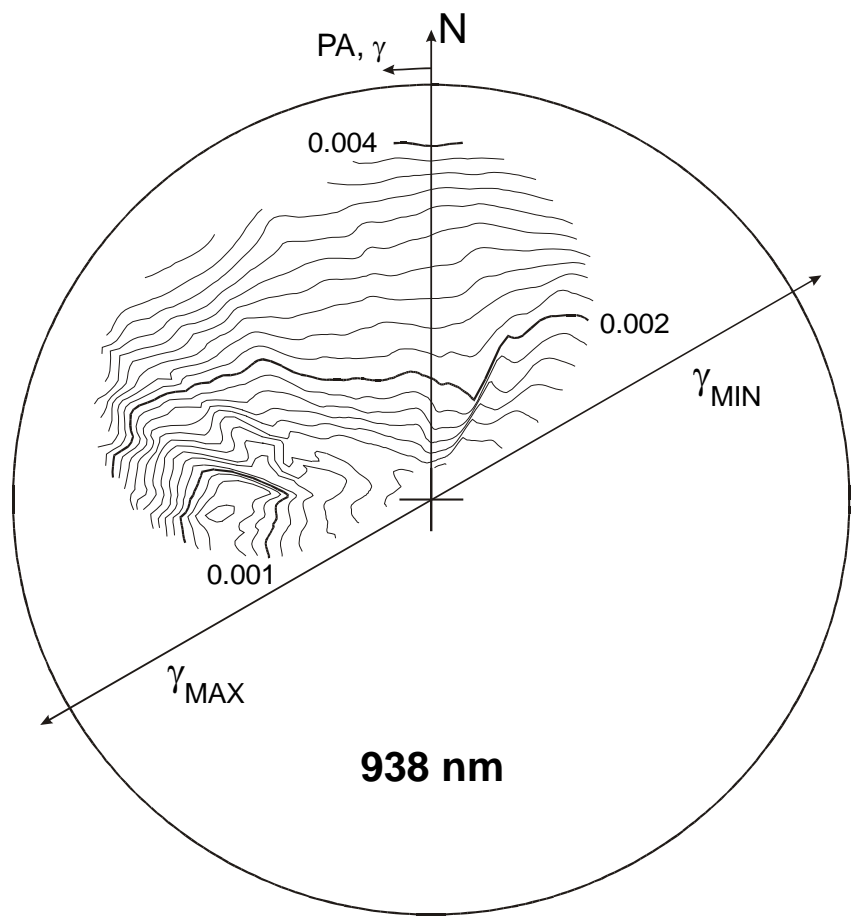

*Figure 2.* Umbra darkening factor distribution for two observational spectral bands with position angles denotations.

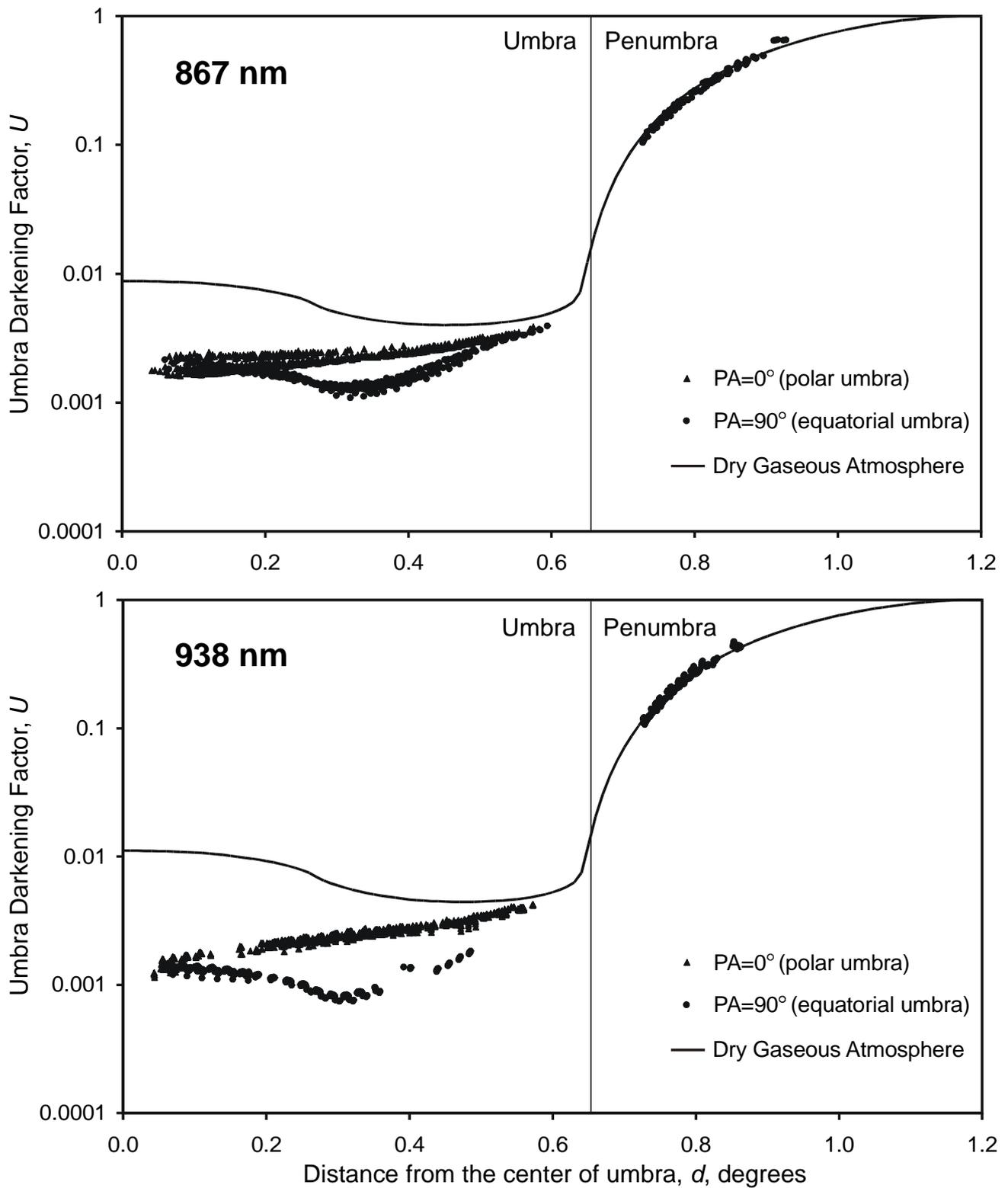

*Figure 3.* *Dependencies of umbra darkening factor on the distance from the center of umbra for two wavelengths and two position angles compared with theoretical curves for dry gaseous atmosphere.*

$$U(d, PA) = \frac{I(d, PA)}{I_0} = \frac{1}{I_0} \int_0^{\rho_0} \int_0^{2\pi} S(\rho) \left( \frac{\pi_0}{\pi_0 - \beta_1} E(\beta_1, \gamma_1) + \frac{\pi_0}{\beta_2 - \pi_0} E(\beta_2, \gamma_2) \right) \rho \, d\varphi \, d\rho,$$

$$I_0 = \int_0^{\rho_0} \int_0^{2\pi} S(\rho) \rho \, d\varphi \, d\rho, \quad \beta_{1,2} = \beta_{1,2}(d, \rho, \varphi); \quad \gamma_{1,2} = PA - \psi_{1,2}(d, \rho, \varphi) \quad (1).$$

Here $I$ and $I_0$ are the brightness values of Sun visible from this point beyond the Earth's atmosphere and outside the eclipse, $S(\rho)$ is the surface brightness distribution of the solar disk, $\pi_0$ is the horizontal parallax of the Moon, $\beta_{1,2}$ and $\gamma_{1,2}$ are the refraction and position angles of the solar disk fragment emission. Other designations can be seen in the Figure 1. Note that the picture visible from the Moon is mirror-transformed relatively the one seen from the Earth (Figure 2), and the angles $PA$ and $\gamma$ in the Figure 1 are counted clockwise. The ray dilution factor $E(\beta, \gamma)$ is expressed as follows [2]:

$$E(\beta, \gamma) = 0, \quad \beta > \beta_0;$$
$$E(\beta, \gamma) = T_G(h(\beta)) \cdot T_A(h(\beta), \gamma) \cdot \frac{1}{1 - L\frac{d\beta}{dh}}, \quad 0 \leq \beta \leq \beta_0;$$
$$E(\beta, \gamma) = 1, \quad \beta < 0 \quad (2).$$

Here $h(\beta)$ is the perigee altitude of the ray path refracting by the angle $\beta$, $L$ is the distance between the Earth and the Moon, $T_G$ is the transparency of the dry gaseous atmosphere along this path, that can be assumed to be independent on position angle $\gamma$, and $T_A$ is the transparency of the atmospheric aerosol along the same path, which depends on $\gamma$. If we observe inside the water vapor absorption band, the term $T_W(h(\beta), \gamma)$ will also appear.

To regularize the solution, we assume (following [2]) that the dependency $T_A(h(\beta), \gamma)$ (or the result of its multiplication with $T_W$ for 938 nm) is partially linear by the angles $\beta$ and $\gamma$:

$$T_A(h(\beta), \gamma) = T_{A2}(\beta_k, \gamma_l)\left(1 - \frac{|\beta - \beta_k|}{\beta_S}\right)\left(1 - \frac{|\gamma - \gamma_l|}{\gamma_S}\right) + T_{A2}(\beta_{k+1}, \gamma_l)\left(1 - \frac{|\beta - \beta_{k+1}|}{\beta_S}\right)\left(1 - \frac{|\gamma - \gamma_l|}{\gamma_S}\right) +$$
$$+ T_{A2}(\beta_k, \gamma_{l+1})\left(1 - \frac{|\beta - \beta_k|}{\beta_S}\right)\left(1 - \frac{|\gamma - \gamma_{l+1}|}{\gamma_S}\right) + T_{A2}(\beta_{k+1}, \gamma_{l+1})\left(1 - \frac{|\beta - \beta_{k+1}|}{\beta_S}\right)\left(1 - \frac{|\gamma - \gamma_{l+1}|}{\gamma_S}\right) \quad (3).$$

Here $\beta_k$ and $\beta_{k+1}$ are the left and right neighbor grid points by the refraction angle, $\gamma_l$ and $\gamma_{l+1}$ are the same for position angle, $\beta_S$ and $\gamma_S$ are the steps of the grid. For the wavelength $\lambda_1$ (867 nm) the integral in formula (1) turns into the sum

$$U(d, PA, \lambda_1) = \sum_k \sum_l U_{kl}(d, PA, \lambda_1) \cdot T_{A2}(\beta_k, \gamma_l) \quad (4),$$

where the coefficients

$$U_{kl}(d, PA, \lambda_1) = \frac{1}{I_0} \int_0^{\rho_0} \int_0^{2\pi} S(\rho, \lambda_1) \cdot [\frac{\pi_0}{\pi_0 - \beta_1} T_G(h(\beta_1), \lambda_1) \frac{1}{1 - L\frac{d\beta_1}{dh}} B_k(\beta_1) G_l(\gamma_1) +$$
$$+ \cdot \frac{\pi_0}{\beta_2 - \pi_0} T_G(h(\beta_2), \lambda_1) \frac{1}{1 - L\frac{d\beta_2}{dh}} B_k(\beta_2) G_l(\gamma_2)] \cdot \rho \, d\varphi \, d\rho \quad (5)$$

can be calculated using the dry gaseous atmosphere model. Here

$$B_k(\beta) = 1 - \frac{|\beta - \beta_k|}{\beta_S}, \quad |\beta - \beta_k| < \beta_S; \quad B_k(\beta) = 0, \quad |\beta - \beta_k| \geq \beta_S;$$

$$G_l(\gamma) = 1 - \frac{|\gamma - \gamma_l|}{\gamma_S}, \quad |\gamma - \gamma_l| < \gamma_S; \quad G_l(\gamma) = 0, \quad |\gamma - \gamma_l| \geq \gamma_S \quad (6).$$

The formula (3) is the linear equation for the values $T_{A2}(\beta_k, \gamma_l)$. Having written it for different values of $d$ and $PA$, we obtain the system that can be solved by the minimum squares method. But solutions can still have large errors due to mathematical incorrectness of the problem. For further regularization we have to take into account all a priori information about the solutions. First of all, the solutions are the values of atmosphere transparency, which has the following property:

$$0 \leq T_{A2}(\beta_k, \gamma_l) \leq 1 \quad (7).$$

The dependency of umbra darkening factor on $d$ (see figure 3) is like the ones for 2004 eclipses and we can follow [2] using the natural assumption

$$T_{A2}(\beta_k, \gamma_l) = 1, \quad \beta_k = 0;$$
$$T_{A2}(\beta_k, \gamma_l) = 0, \quad \beta_k \geq 1° \quad (8).$$

Here we mean that the atmosphere is transparent at its upper border and absorbs all tangent emission propagating through its lowest layers (less than 0.7 km). Moreover, for the better accuracy of the results at the edges of the grid we will calculate them in the range of angles $\gamma$ between $\gamma_{MIN} = -60°$ and $\gamma_{MAX} = 120°$ corresponding to minimal and maximal $PA$ values (as shown in the Figure 2) assuming

$$T_{A2}(\beta_k, \gamma) = T_{A2}(\beta_k, \gamma_{MIN}), \quad \gamma < \gamma_{MIN};$$
$$T_{A2}(\beta_k, \gamma) = T_{A2}(\beta_k, \gamma_{MAX}), \quad \gamma > \gamma_{MAX} \quad (9).$$

Following [2], resolution by the refraction angle $\beta_S$ is accepted to be 0.2°, that is a little bit lower than the angular radius of the Sun. It corresponds to the altitude resolution in the troposphere 3-4 km. Better resolution leads to errors increase in solution. The corresponding resolution by the position angle $\gamma_S$ is equal to 15°. So we have six grid points by angle $\beta$ (from 0.0° to 1.0° through 0.2°) and 13 grid points by angle $\gamma$ (from –60° to 120° through 15°). The number of independent parameters is less than 78 due to a priori information.

Having run this procedure for 867 nm observational data, we obtain the array $T_{A2}(\beta_k, \gamma_l)$. According to slow wavelength dependency of refraction and aerosol extinction alongside with small difference of two observational wavelengths compared with the wavelengths themselves, we assume these parameters to be the same for $\lambda_2$ equal to 938 nm. In order to maximize a priori data for this wavelengths the values $T_{A2}(\beta_k, \gamma_l)$ are included to the matrix. Thus, the equations system for the second wavelength takes the form

$$U(d, PA, \lambda_2) = \sum_k \sum_l [T_{A2}(\beta_k, \gamma_l) \cdot U_{kl}(d, PA, \lambda_2)] \cdot T_{W2}(\beta_k, \gamma_l) \quad (10).$$

The number of independent parameters here will be less than in the equations (4) for 867 nm, since the pairs $(k, l)$ for which $T_{A2}(\beta_k, \gamma_l) = 0$ will be excluded; $T_W$ for these $k$ and $l$ can not be found. It is basically the case of low altitudes where the atmospheric aerosol absorbs all tangent emission. But if $T_{A2}(\beta_k, \gamma_l)$ is close to 1, corresponding $T_W$ values can be found with higher accuracy.

The same procedure consequence is used to run one-dimensional reintegration and to find the arc-average values of $T_{A1}(\beta_k, PA)$ and $T_{W1}(\beta_k, PA)$ for different $PA$ values in the same range as for $\gamma$ angles by the method described in [2].

## 4. Results.

Figures 4 and 5 show the results of one- and two-dimensional reintegration procedures of calculation of $T_A$ values for 867 nm at different perigee altitudes (or $\beta$ angles) depending on the position angles $PA$ (for one-dimensional procedure) or $\gamma$ (for two-dimensional procedure). The coordinates of the Earth's limb points corresponding to the position angles are also shown. One-dimensional reintegration results are shown by the connected dots, two dimensional results are shown by the broken lines according to formula (3), minimum square method errors bars are shown for the points where $0<T_{A2}<1$. It is clear to see the good agreement of one- and two-dimensional analyses for the refraction angles 0.2° and 0.4° corresponding to ray perigee altitudes 14.9 and 10.5 km, respectively (Figure 4). It draws the altitude range in the troposphere where the method of lunar eclipse photometry is effective to detect the absorbing components. Upper troposphere (14.9 km) is basically free from aerosol absorption except northern polar regions and central Canada. Aerosol at this altitude was not detected anywhere during 2004 lunar eclipses [2]. It also was not detected by twilight polarization measurements in Crimea from 2000 until October 2006 [5-7], but appeared there in December, 2006. This aerosol is possible part of polar stratospheric clouds moved southwards by polar stratospheric vortex.

Atmospheric aerosol at the altitude 10.5 km appears near the equator and polar regions, disappearing in the tropical zone. The same was observed during the eclipse of May, 4th, 2004. Aerosol absorption at 10.5 km is practically absent over the Siberia and Himalayan mountains, that is natural for the continental winter conditions.

Analysis of aerosol extinction for larger refraction angles 0.6° and 0.8° and corresponding ray perigee altitudes 6.2 and 3.2 km meets serious difficulties (see Figure 5). Radiation propagating along such path in the atmosphere transfers to the deep umbra regions. Being observed from there, the Sun is hidden deep behind the Earth and its disk covers wide range of $\psi$ angles (see Figure 1). Two-dimensional reintegration data with resolution 15° show rapid variations with larger errors, averaged one-dimensional reintegration data contain no remarkable features. The only thing that should be noted from two-dimensional analysis is vanishing of $T_{A2}$ values in the equatorial and northern tropical latitudes. It was expected owing to huge aerosol concentration near the equator and light obscuration by Himalayan mountains in the tropics.

The aerosol data obtained from the observations at the wavelength 867 nm and described above is then used to find the water vapor transparency $T_{W2}(\beta_k, \gamma_l)$ using the 938 nm data by the solution of the equations system (10). For the altitudes 3.2 and 6.2 km, where the parameters $T_{A2}(\beta_k,\gamma_l)$ are small and found with bad accuracy, the numbers $T_{W2}$ will also have bad accuracy or even can not be found (if corresponding $T_{A2}(\beta_k, \gamma_l)$ value is equal to zero). We will focus our attention on the results for the ray perigee altitudes 10.5 and 14.9 km.

Figure 6 shows the data of one-dimensional and two-dimensional reintegration for these perigee altitudes, the designations are analogous to the Figure 4. The most remarkable feature is strong water vapor light absorption at the altitude 10.5 km near the equator. The same figure contains the distribution of total water vapor column density along the limb obtained by the AMC-DOAS technique in SCIAMACHY space mission for the same day. The technique is described in [8-10] and updated in [11]. Good anti-correlation of one-dimension reintegrated value $T_{W1}(0.4°, PA)$ for perigee altitude $h=10.5$ km and total column amount $C_0$ is clear to see. Optical depth of the water vapor along this tangent path is related with column density as

$$\tau_h = -\ln T_{W1}(\beta(h), PA) = D \cdot C_0 \quad (11),$$

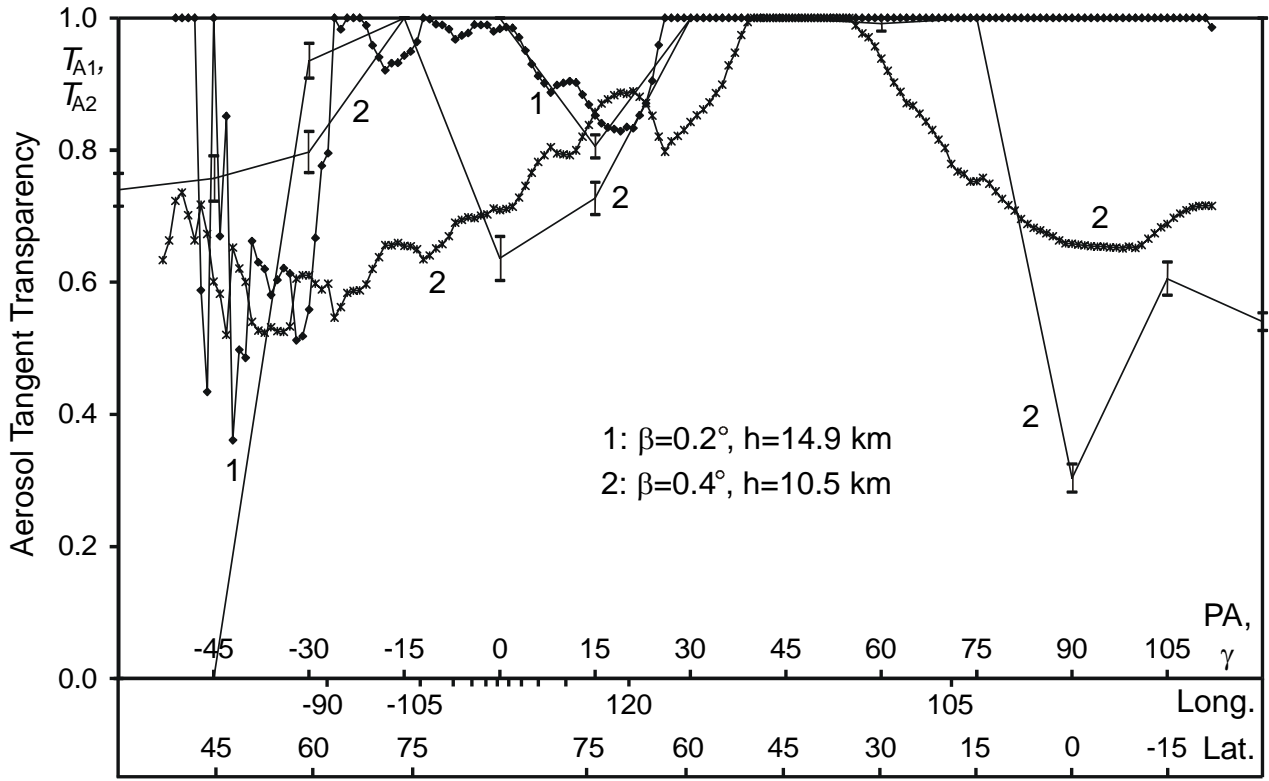

*Figure 4.* Atmospheric aerosol tangent transparency for different locations at the limb in the upper troposphere as the result of one-dimensional (connected dots) and two-dimensional (broken lines) reintegration procedures.

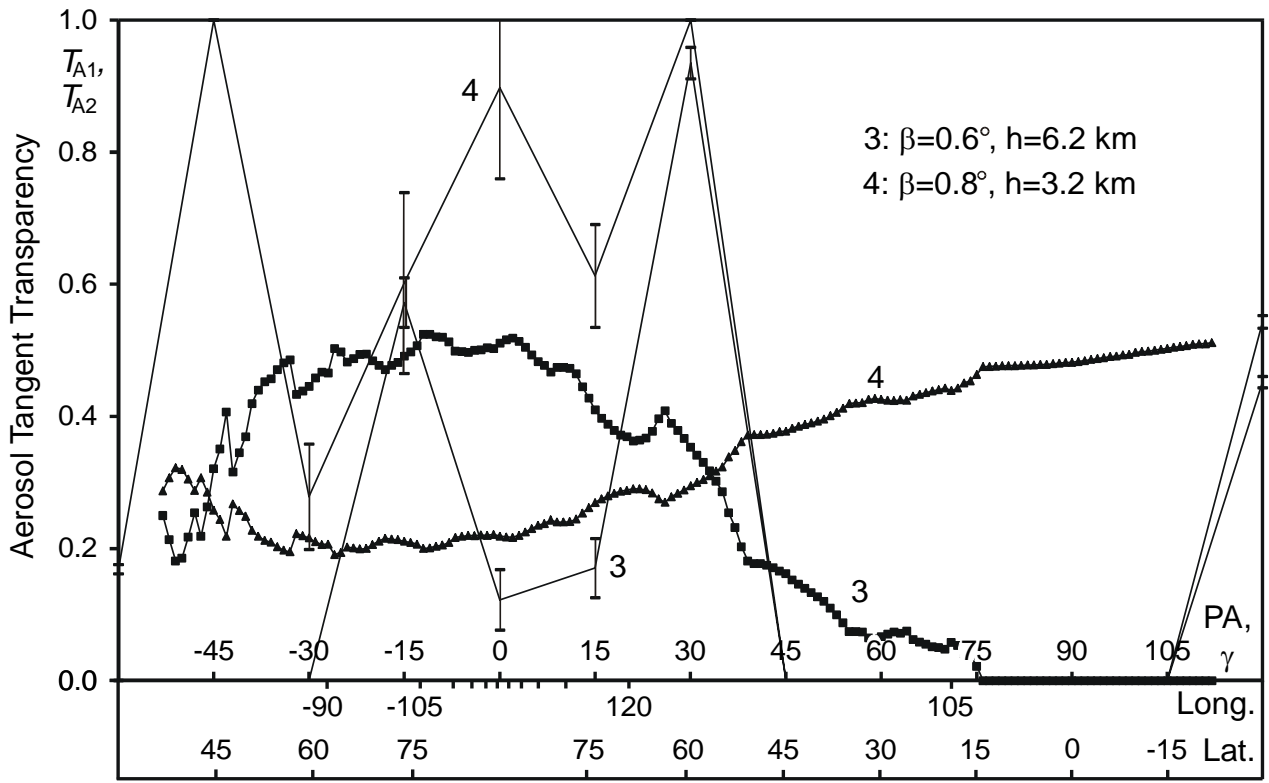

*Figure 5.* Atmospheric aerosol tangent transparency for different locations at the limb in the lower troposphere as the result of one-dimensional (connected dots) and two-dimensional (broken lines) reintegration procedures.

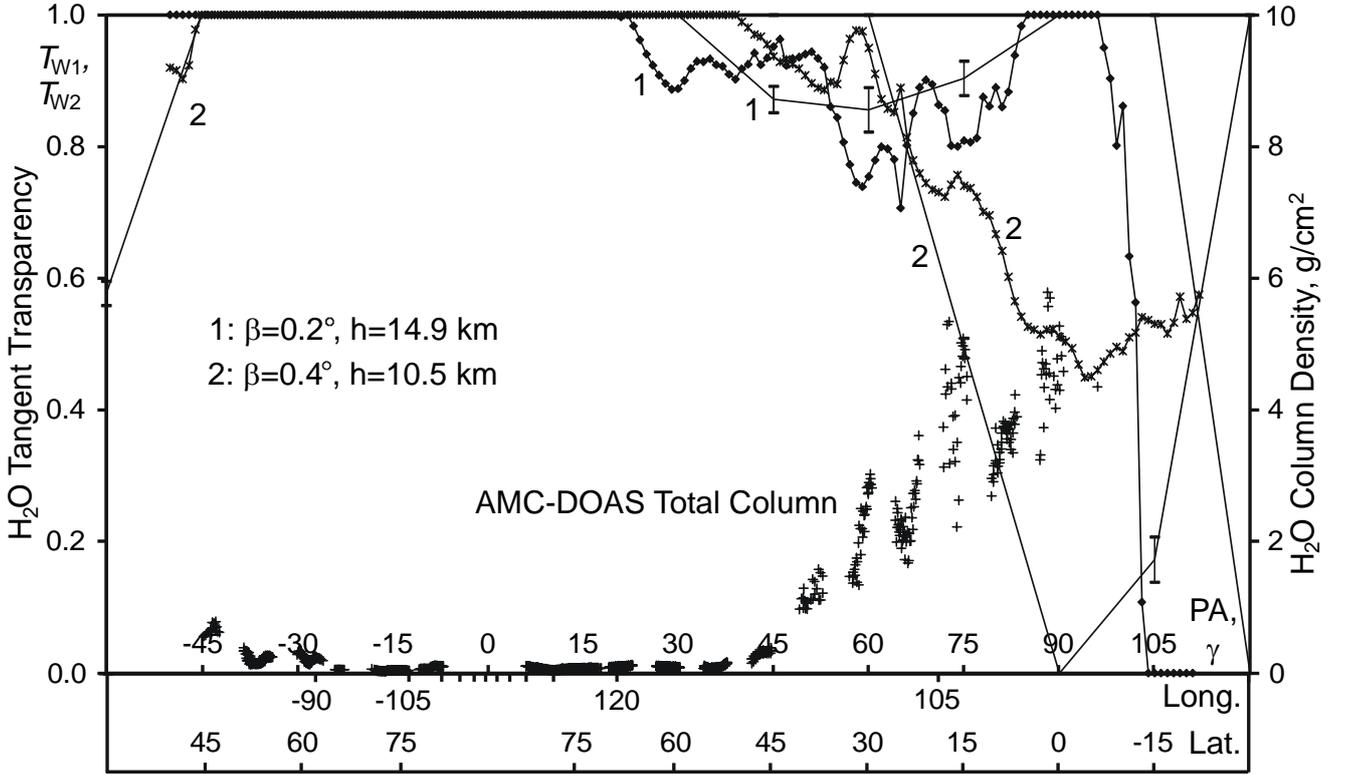

*Figure 6.* Water vapor tangent transparency for different locations at the limb in the upper troposphere as the result of one-dimensional (connected dots) and two-dimensional (broken lines) reintegration procedures compared with AMC-DOAS water vapor total column density.

where the coefficient $D$ is equal to 0.12 cm$^2$/g, being quite stable along the meridian. Assuming the altitude distribution of the water vapor to be exponential with the scale $H$ and small value of refraction angle, we write the equation for water vapor optical depth along the tangent path with perigee altitude $h$:

$$\tau_h = \int_{-\infty}^{\infty} \sigma \cdot n_0 \cdot e^{-\frac{\sqrt{(R+h)^2+x^2}-R}{H}} dx \approx \sigma \cdot n_0 \cdot e^{-\frac{h}{H}} \int_{-\infty}^{\infty} e^{-\frac{x^2}{2RH}} dx = \sigma \cdot n_h \sqrt{2\pi RH} \qquad (12).$$

Here $R$ is the radius of the Earth, $\sigma$ is the effective water vapor cross section in the observational passband, $n_0$ and $n_h$ are the vapor concentration values near the ground and at the altitude $h$. The water vapor column density above the altitude $h$ is equal to

$$C_h = n_h mH = \frac{\tau_h m}{\sigma}\sqrt{\frac{H}{2\pi R}} \qquad (13).$$

Here $m$ is the water molecule mass. Substituting (11) into (13) and taking into account the exponential altitude dependence of water vapor amount, we obtain:

$$\frac{C_h}{C_0} = \frac{mD}{\sigma}\sqrt{\frac{H}{2\pi R}} = e^{-\frac{h}{H}} \qquad (14).$$

Solving this equation numerically, we obtain the values of $H$ (1.3 km) and $C_h/C_0$ (about $4\cdot10^{-4}$). It shows the rapid decrease of water vapor amount with the altitude. The scale weakly depends on the latitude. For the equatorial troposphere, assuming $T_W$ to be equal to 0.5 (one-dimensional reintegration result), we obtain the water vapor concentration value for 10.5 km using formula (12): $5 \cdot 10^{14}$ cm$^{-3}$. For the ray path with the perigee altitude equal to 14.9 km weak traces of water vapor (concentration estimation for the same $H$ is about $10^{14}$ cm$^{-3}$) are seen northwards from equator. It is also seen as rapid fall of $T_W$ southwards from equator, but numerical analysis at the edge of observed range of $PA$ and $\gamma$ can lead to sufficient errors.

Summarizing the results, Figure 7 shows the distribution of one-dimension reintegration values of tangent path extinction $T_{A1}$ and $T_{W1}$ for the perigee altitude 10.5 km compared with the weather map provided by Space Science and Engineering Center, University of Wisconsin-Madison, for the March, 4, 2007, 6h UT (a few hours after the eclipse).

## 5. Discussion and conclusion

The paper describes two IR-bands photometric analysis of total lunar eclipse of March, 3, 2007 and its application to the retrieval of latitude and altitude distribution of aerosol and water vapor in the troposphere. From the one hand, the lunar eclipse gives the unique possibility to hold such type of remote sensing from the ground. From the other hand, necessity of reverse problem solution with account of large angular diameter of the Sun restricts the resolution of the method on the position angle (or latitude) and altitude. Analysis made in this paper shows that good results can be achieved in upper troposphere layers, from 10 km to tropopause.

Aerosol distribution is quite similar to the one obtained during two lunar eclipses in 2004 [2], that is natural since the visual brightness of 2007 eclipse was close to the 2004 ones. It does not follow the hypothesis of Danjon described in [1] predicting the decrease of eclipse brightness and corresponding increase of aerosol level in the upper troposphere after the solar activity minimum. The exception is the central and northern Canada, where the aerosol at the altitude 14.9 km absent anywhere in 2004 was detected alongside with enchanted level of aerosol at the altitude 10.5 km. Siberian and Himalayan regions of the Earth's limb are free from aerosol in the upper troposphere. Such aerosol appears in the near-equatorial part of South-Eastern Asia.

Analysis of aerosol distribution below 10 km is difficult owing to strong tangent rays absorption and loss of resolution by position angle cased by the angular size of the Sun. The only effects that are clearly seen are the increase of aerosol extinction near the equator and obscuration by Himalayan mountains in the tropics.

The data obtained in the second observational passband in the range of water vapor absorption give the possibility to investigate the distribution of water vapor at the same altitudes above the limb. The 10.5 km results show the good agreement between one- and two-dimensional models of reintegration and correlation with total water vapor column densities obtained by SCIAMACHY AMC-DOAS technique [8-11]. Analysis of this correlation gives the possibility to calculate the character scale of water vapor altitude distribution (1.3 km), which is sufficiently less than the one for other atmosphere components, and to determine the water vapor concentration in the upper troposphere above the limb.

**Acknowledgements**

Authors are thankful to Stefan Noël (Institute of Environmental Physics/Remote Sensing, University of Bremen) for supplying the SCIAMACHY AMC-DOAS water vapor data and to Space Science and Engineering Center, University of Wisconsin-Madison for the weather map for the eclipse date. We would also like to thank V.I. Shenavrin (Sternberg Astronomical Institute) for the help during the observations, N.N. Shakhvorostova (Astro-Space Center, Lebedev's Physical Institute of RAS), K.B. Moiseenko (Institute of Atmospheric Physics of RAS), A.M. Feigin (Institute of Applied Physics of RAS) and E. Tsimerinov for some useful remarks.

O.S. Ugolnikov is supported by Russian Science Support Foundation grant.

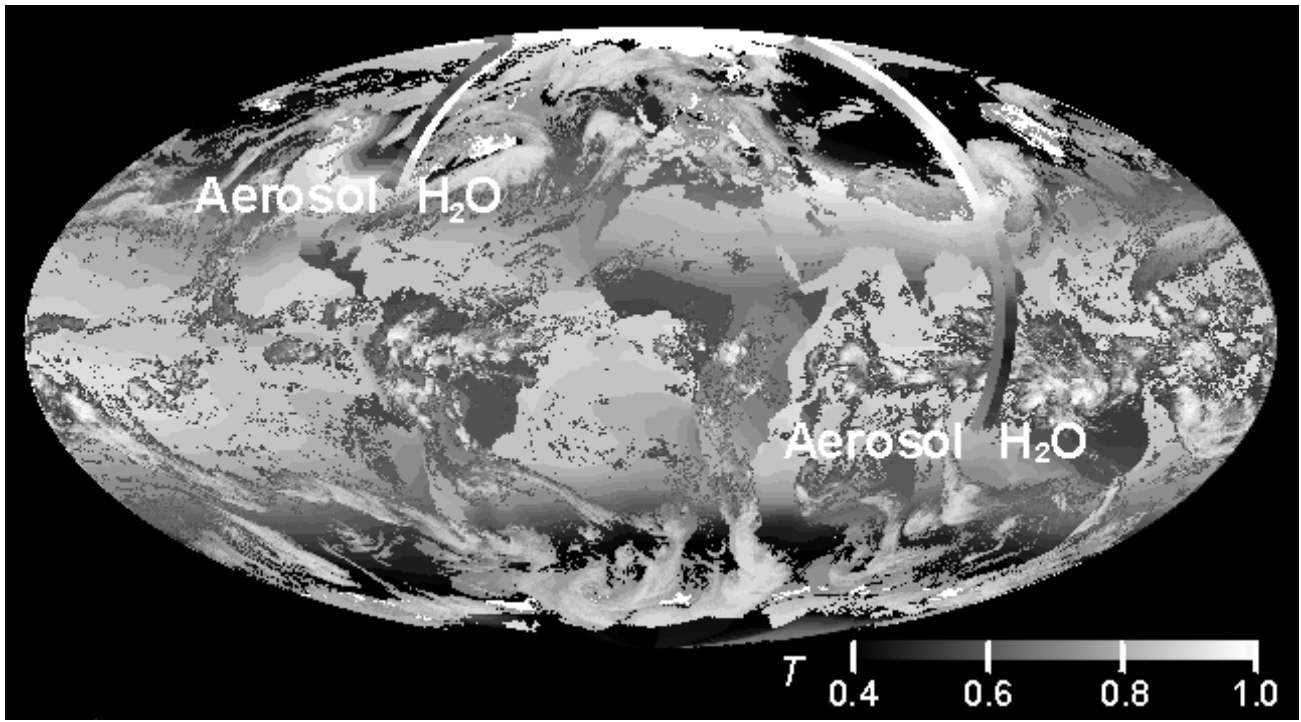

*Figure 7.* *Transparency of atmospheric aerosol and water vapor by the tangent path with perigee altitude 10.5 km above the different locations on the limb compared with weather map provided by Space Science and Engineering Center, University of Wisconsin-Madison.*